\documentclass[%
 aip,
 amsmath,amssymb,
 reprint,%
]{revtex4-1}

\usepackage{graphicx}
\usepackage{dcolumn}
\usepackage{bm}

\usepackage[utf8]{inputenc}
\usepackage[T1]{fontenc}
\usepackage{mathptmx}
\usepackage{etoolbox}

\makeatletter
\def\@email#1#2{%
 \endgroup
 \patchcmd{\titleblock@produce}
  {\frontmatter@RRAPformat}
  {\frontmatter@RRAPformat{\produce@RRAP{*#1\href{mailto:#2}{#2}}}\frontmatter@RRAPformat}
  {}{}
}%
\makeatother
\begin{document}

\preprint{AIP/123-QED}

\title[Optimizing Topological Switching in Confined 2D-Xene Nanoribbons via Finite-Size Effects]{Optimizing Topological Switching in Confined 2D-Xene Nanoribbons via Finite-Size Effects}
\author{Muhammad Nadeem}
\email{mnadeem@uow.edu.au, michael.fuhrer@monash.edu, xiaolin@uow.edu.au}
\altaffiliation{ARC Centre of Excellence in Future Low-Energy Electronics Technologies (FLEET), University of Wollongong, Wollongong, New South Wales 2525, Australia}
\affiliation{Institute for Superconducting and Electronic Materials (ISEM), Australian Institute for Innovative Materials (AIIM), University of Wollongong, Wollongong, New South Wales 2525, Australia.}
\author{Chao Zhang}
\affiliation{School of Physics, University of Wollongong, Wollongong, NSW 2522, Australia}
\author{Dimitrie Culcer}
\altaffiliation{ARC Centre of Excellence in Future Low-Energy Electronics Technologies (FLEET), University of New South Wales, Sydney 2052, Australia.}
\affiliation{School of Physics, University of New South Wales, Sydney 2052, Australia.}
\author{Alex R. Hamilton}
\altaffiliation{ARC Centre of Excellence in Future Low-Energy Electronics Technologies (FLEET), University of New South Wales, Sydney 2052, Australia.}
\affiliation{School of Physics, University of New South Wales, Sydney 2052, Australia.}
\author{Michael S. Fuhrer}
\altaffiliation{ARC Centre of Excellence in Future Low-Energy Electronics Technologies (FLEET), Monash University, Clayton, Victoria 3800, Australia.}
\affiliation{School of Physics and Astronomy, Monash University, Clayton, Victoria 3800, Australia.}
\author{Xiaolin Wang}
\altaffiliation{ARC Centre of Excellence in Future Low-Energy Electronics Technologies (FLEET), University of Wollongong, Wollongong, New South Wales 2525, Australia}
\affiliation{Institute for Superconducting and Electronic Materials (ISEM), Australian Institute for Innovative Materials (AIIM), University of Wollongong, Wollongong, New South Wales 2525, Australia.}

\begin{abstract}
In a blueprint for topological electronics, edge state transport in a topological insulator material can be controlled by employing a gate-induced topological quantum phase transition. Here, by studying the width dependence of electronic properties, it is inferred that zigzag-Xene nanoribbons are promising materials for topological electronics with a display of unique physical characteristics associated with the intrinsic band topology and the finite-size effects on gate-induced topological switching. First, due to intertwining with intrinsic band topology-driven energy-zero modes in the pristine case, spin-filtered chiral edge states in zigzag-Xene nanoribbons remain gapless and protected against backward scattering even with finite inter-edge overlapping in ultra-narrow ribbons, i.e., a 2D quantum spin Hall material turns into a 1D topological metal. Second, mainly due to width- and momentum-dependent tunability of the gate-induced inter-edge coupling, the threshold-voltage required for switching between gapless and gapped edge states reduces as the width decreases, without any fundamental lower bound. Third, when the width of zigzag-Xene nanoribbons is smaller than a critical limit, topological switching between edge states can be attained without bulk bandgap closing and reopening. This is primarily due to the quantum confinement effect on the bulk band spectrum which increases the nontrivial bulk bandgap with decrease in width. The existence of such protected gapless edge states and reduction in threshold-voltage accompanied by enhancement in the bulk bandgap overturns the general wisdom of utilizing narrow-gap and wide channel materials for reducing the threshold-voltage in a standard field effect transistor analysis and paves the way toward low-voltage topological devices.
\end{abstract}

\maketitle
\section{\label{intro}Introduction}
Two-dimensional topological insulators are promising materials for topological quantum electronic devices where edge state transport can be controlled by a gate-induced electric field \cite{WaryL,Topo-Elec}. In general, edge state transport can be controlled either by a perpendicular electric field, which drives a topological phase transition via bulk bandgap closing and reopening \cite{Ezaw13APL, Liu14-natmat, Liu15-nano, pan15sci, Qian14, zhange-natnano17,Molle-natmat17, Collins18, Muhd-nano} or via inter-edge tunneling between gapped edge states, assisted by a transverse electric field \cite{TTFET}. In the latter case, though a very weak transverse electric field is sufficient to induce inter-edge tunneling, edge state conductance quantization may remain a challenge constraining the geometry of topological insulator ribbons \cite{TTFET}. In the former case, a blueprint for topological quantum electronics, the strength of the critical electric field required for topological switching depends upon the strength of quantum mechanical perturbations, such as the spin-orbit interaction (SOI) \cite{Kane05a,Kane05b} and Bernevig–Hughes–Zhang (BHZ) mass term \cite{Bernevig06,Konig08}, which reflect the bulk band topology and therefore lead to a quantized edge state conductance. In this class, numerous theoretical proposals for 2D topological insulator materials, which exhibit electrically-driven topological phase transition, such as staggered sublattice potentials \cite{Ezaw13APL,Molle-natmat17, Muhd-nano}, mirror symmetry breaking \cite{Liu14-natmat}, and the Stark effect \cite{Qian14,Liu15-nano, pan15sci, zhange-natnano17, Collins18}, have been put forward. Among these various proposals, electric field switching has been demonstrated \cite{Collins18} in ultrathin (monolayer or bilayer) Na\textsubscript{3}Bi where the experimentally reported bandgap of 360 meV significantly exceeds the thermal energy at room temperature (25 meV) and the critical electric field is about 1.1 V nm$^{-1}$.\par

Though a large topological bulk bandgap is desirable to enable quantum spin Hall (QSH) phenomena at room temperature, one of the main challenges with such blueprint topological switching mechanism is the requirement of unrealistically large critical electric field to close the topological bandgap \cite{Molle-natmat17, Pseudo-TFET-HgTe}. For instance, the critical electric field is of the order of 0.05 V nm$^{-1}$ for silicene \cite{Molle-natmat17}, 1.0 V nm$^{-1}$ for stanene \cite{Molle-natmat17}, and 1.42 V nm$^{-1}$ for 1T$^\prime$-MoS$_2$ \cite{Qian14}. The critical electric field further increases for heavy elemental topological insulators such as Bi/SiC \cite{Reis17} where the experimentally reported bandgap is 800 meV. In light of this, only recently, it is shown that the critical electric field can be significantly reduced via the tunable Rashba effect in 2D-Xenes (G, Si, Ge, Sn, and P, As, Sb, Bi) \cite{Muhd-nano}. Though the Rahsba effect is considerably large in heavy elemental 2D-Xenes such as functionalized bismuthene, the Rashba effect remains negligibly small for relatively lighter group-IV (Si, Ge) and group-V (P, As) elemental 2D-Xenes \cite{Muhd-nano}.\par

Here we note that, apart from the relativistic quantum mechanical phenomenon of SOI which plays a central role in characterizing the band topology and limiting the critical electric field, the finite-size geometry incorporates two additional critical phenomena: quantum confinement effects on the bulk subbands and inter-edge coupling between spin-filtered chiral edge states. For the study of fundamental phenomena in both laboratory and device applications, it is crucial to investigate the fundamental topological features and the edge state transport in the finite-size geometry of a topological insulator. Finite-size effects have been studied for various topological insulator materials via the thickness dependence of surface electronic dispersion in thin films of 3D topological insulators \cite{Shan10,SCZhnag,Lu-Shan10} and Dirac semimetals \cite{pan15sci, Collins18} and the width dependence of edge state electronic dispersion in 2D topological insulator materials such as HgTe \cite{Zhou08}, transition metal dichalcogenides (TMDC) with 1T$^\prime$ phase, TMDC-1T$^\prime$ \cite{Das20}, and 2D-Xenes \cite{XNRs-Ezawa, XNRs-Kim, XNRs-Son, XNRs-Brey2, XNRs-Ezawa2, XNRs-Rubby}. However, less attention has been devoted to finite-size effects on the gate-induced topological switching, a central part for the working of topological electronics devices.\par

By studying the width dependence of electronic properties in zigzag Xene nanoribbons (ZXNRs), it is demonstrated that both the SOI-induced barrier in the bulk and the gate-control of quantized conductance along the edges can be optimized via finite-size effects. It is inferred that ZXNRs are promising materials for topological electronics with a display of unique physical characteristics associated with the intrinsic band topology and the finite-size effects on gate-induced topological switching. Through tight binding calculations of band dispersion, density of states (DOS), conductance quantization, edge state wave functions and their width dependence, we highlight several results that are crucial for understanding fundamental aspects and developing novel device concepts. Our findings and the analysis presented for ZXNRs remain valid for any 2D topological insulator material with buckled honeycomb structure terminated on zigzag edges.\par

First, spin-filtered chiral edge states in ZXNRs remain gapless and protected against backward scattering even with finite inter-edge overlapping in ultra-narrow ribbons, i.e., a 2D quantum spin Hall (QSH) material turns into 1D topological metal. Such robustness in ZXNRs is deeply rooted in intertwining between SOI-induced spin-filtered chiral modes and the intrinsic band topology-driven energy-zero modes in pristine honeycomb lattice structures. Furthermore, the topological protection of 1D metallic modes is a consequence of different momentum space locations for edge state crossings (at time-reversal invariant momenta (TRIM) $k=\pi$) and anti-crossings (around valleys k=K/K$^\prime$). Here the edge state crossing point is a Dirac point formed by edge state gapless Dirac dispersion while the edge state anti-crossing point is a momentum space location where the edge state spectrum becomes a massive Dirac dispersion. This is highly contrasting from other 2D topological insulator materials with inverted band structure, in which the edge state crossing and anti-crossing points coexist, and in which hybridization due to inter-edge overlapping opens an energy gap and leads to a gapped edge state spectrum \cite{Zhou08}. For instance, in inverted band topological insulators such as thin films of X$_2$Y$_3$ [X=Bi,Sb;Y=Te,Se] \cite{Shan10,SCZhnag,Yi} semiconductors, type-I HgTe/CdTe \cite{Bernevig06, Konig07} and type-II InAs/GaSb/AlSb \cite{Liu-typeII, Knez-TypeII,Du-TypeII} semiconducting quantum well structures, Na$_3$Bi thin films \cite{pan15sci,Collins18}, and monolayers of TMDC with 1T$^\prime$ phase \cite{Qian14,Wu}, both edge state crossing and anti-crossing points coexist at $\Gamma$-point.\par

Second, the critical electric-field required for switching between gapless and gapped edge states reduces as the width of ZXNRs decreases, without any fundamental lower bound. We demonstrate explicitly that such size dependence of the threshold-voltage stems from a series of non-trivial quantum mechanical phenomena associated with the geometric structure of ZXNRs. First, the edge state wave functions at the crossing point are independent of the edge termination and hence remain insensitive to electric fields. On the other hand, edge state wave functions and the gate-induced coupling between overlapping edge states across anti-crossing points are strongly dependent on the particular edge termination and hence can be tuned via a gate electric field. Second, with a decrease in width, the momentum space location of the edge state anti-crossing points moves away from the valleys toward the TRIM ($k=\pi$). Furthermore, at particular momenta around the anti-crossing points, the magnitude of the inter-edge overlap increases with decrease in width. As a result, gate-induced coupling between spin-filtered chiral edge states is enhanced as the ZXNR width is reduced. It shows that finite-size effects on the edge spectrum play a central role in optimizing the gate-controlled edge state transport, such that size dependence of the threshold-voltage stems from width- and momentum-dependent tunability of the gate-induced coupling between inter-edge spin-filtered chiral states.\par

Third, when the width of ZXNRs is smaller than a critical limit, quantum confinement enhances the topological bulk bandgap and, hence, the energy spacing between the bulk subbands and the edge states, which in turn leads to topological switching between gapless and gapped edge states without bulk bandgap closing. Both of these finite-size phenomena, central to the control of edge state transport, are completely missing in wide ZXNRs: there  the critical electric-field is limited by the SOI-induced barrier, and the topological phase transition is accompanied by bulk bandgap closing and reopening.\par

It is important to note that the threshold reduction could in principle be achieved with a built-in electric field due to static charges. However, for a fixed SOI in ZXNRs, a simple enhancement of the built-in electric field also reduces the topological bandgap in the QSH phase, which may be detrimental to dissipationless quantized edge state conductance due to admixing of edge modes with bulk subbands. In this regard, the size-dependent and momentum-dependent tunability of gate-induced inter-edge coupling is a novel mechanism that reduces the critical gate electric field even as the topological bulk bandgap is enhanced by quantum confinement. On the one hand, it reduces the threshold-voltage by lowering the SOI-induced barrier in the bulk, on the other hand it enhances the bulk bandgap even in lighter monoelemental 2D-Xenes such that the detrimental contributions from the bulk material to the edge current are avoided and allows safe residing of the chemical potential within this gap. \par 

\begin{figure*}
\includegraphics[scale=0.9]{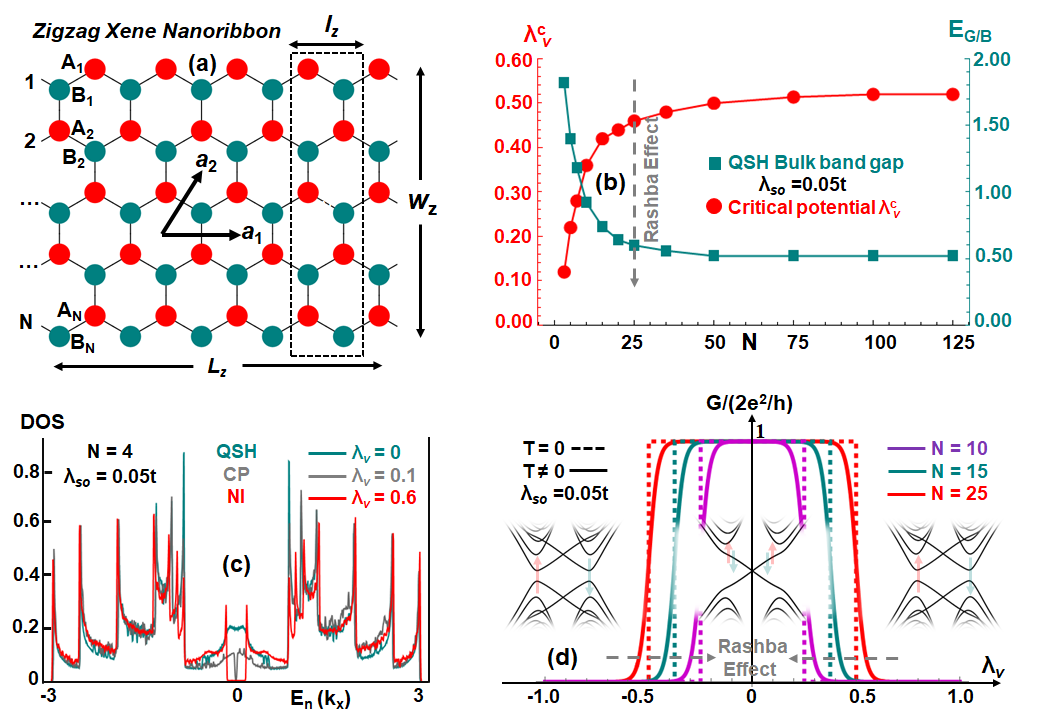}
\caption{\label{TQFET-FSE}\textbf{Optimization of electronic properties in quantum confined ZXNRs}. \textbf{(a)} A ZXNR with lattice parameters. \textbf{(b)} Width dependence of threshold-voltage and bulk bandgap. With decrease in the width of ZXNRs, threshold-voltage decreases while nontrivial bulk bandgap increases. \textbf{(c)} Density of states and topological phase transition in an ultra-narrow ZXNR with N = 4. In the absence of gate electric field ($\lambda_v=0$), finite density of states (cyan) at zero-energy represent the presence of protected helical edge states in QSH phase. With increasing gate electric field, ZXNR enters into normal insulating (NI) phase (red) with gapped edge states while passing through a critical gapless phase (grey). \textbf{(d)} Quantized edge state conductance and the critical gate electric field for N = 10, 15, and 25. Here, Rashba SOI is ignored for simplicity. When Rashba SOI is incorporated, topological quantum field effect further reduces the threshold-voltage and bulk bandgap. Here N represents the number of zigzag chains and simulates the width of ZXNRs as $W_z=\sqrt{3}Na_0/2$.}
\end{figure*}

These features make quantum confined spin-orbit coupled ZXNRs special for topological quantum devices, enabling optimal gate-controlled transport through edge state channels via finite-size effects on the electronic properties. The existence of gapless edge states and reduction in threshold-voltage accompanied by enhancement in the bulk bandgap overturns the general wisdom of utilizing narrow gap and wide channel materials for reducing the threshold-voltage in a standard field effect transistor analysis, other than negative capacitance mechanism \cite{Dass-NCFET}. Furthermore, the advantage of utilizing ultra-narrow ZXNRs is multi-fold: (i) the availability of large edge state conducting modes for enhanced signal-to-noise ratio via multiple edge state channels, (ii) optimized geometry for topological electronic devices where an array of ZXNRs, set apart by trivial insulating layers/wires along vertical/lateral direction, is sandwiched between top and bottom gates separated by top and bottom dielectrics, and (iii) low-voltage and energy-efficient switching due to compressible subthreshold swing in ZXNRs via topological quantum field effect \cite{Muhd-nano} .\par

\section{\label{TBM}Finite-size effects on ZXNRs}
Figure \ref{TQFET-FSE}(a) shows a ZXNR where the primitive lattice vectors are represented by $a_1=a_0(1,0)$ and $a_2=a_0(1/2,\sqrt{3}/2)$, $d_z$ represents the buckling length, while dashed rectangle (composing dimer line of A and B sublattice sites) represents the unit cell for ZXNR. The width of ZXNR is represented by the length of unit cell (dimer line), $W_z=\sqrt{3}N_da_0/4$ where $N_d=2N$ represents the number of sites in the dimer line and $N$ represents the number of zigzag lines. The length of ZXNR $L_z=Dl_z$, where D represents the number of dimer lines and $l_z$ is the width of a dimer line, can be written as a function of the number of sites in the zigzag line ($N_z$) as $L_z=N_za_0/2$.\par

After the seminal work by Kane-Mele \cite{Kane05a,Kane05b} on graphene, it has been shown that other 2D-Xene nanoribbons (Si, Ge, Sn, and P, As, Sb, Bi) with honeycomb lattice structure are also QSH insulators \cite{HongkiMin06,LiuPRL11, LiuPRB11,XuPRL13,Hsu15,Reis17,Li18}. Among 2D topological insulator materials, quantum spin Hall (QSH) insulators with honeycomb lattice structure terminated on zigzag edges are a special class where spin-filtered chiral modes are intertwined with the intrinsic band topology of the pristine honeycomb lattice structure. In ZXNRs, the intrinsic band topology, characterized by a non-vanishing winding number, leads to energy-zero flat bands in the nontrivial regime of the first Brillouin zone. The sublattice-resolved intrinsic SOI, modeled through next-nearest hopping \cite{Kane05a,Kane05b}, disperses these localized modes into spin-filtered chiral edge states. To simulate the energy-zero modes in the pristine case, spin-filtered chiral edge states in the spin-orbit coupled case, gate-induced topological switching, and the dependence of electronic properties on the width of ZXNRs, in general, we study the next-nearest neighbor tight-binding model Hamiltonian \cite{Kane05a, Kane05b}
\begin{widetext}
\begin{equation}
H = t\sum_{\langle ij\rangle\alpha}c_{i\alpha}^\dagger c_{j\alpha} + i\lambda_{so}\sum_{\langle\langle ij \rangle\rangle\alpha\beta} v_{ij} c_{i\alpha}^\dagger s_{\alpha\beta}^z c_{j\beta} +\frac{\lambda_{v}}{2}\sum_{i\alpha} c_{i\alpha}^\dagger v_{i} c_{i\alpha} + i\lambda_{R}(E_{z})\sum_{\langle ij\rangle\alpha\beta} c_{i\alpha}^\dagger (\textbf{s}_{\alpha\beta} \times \hat{\textbf{d}}_{ij})_{z} c_{j\beta}\;. \label{eq:wideeq}
\end{equation}
\end{widetext}
where the first term is the nearest neighbor hopping generating Dirac dispersion in the vicinity of valleys K(K$^\prime$) while the second term is the intrinsic Kane-Mele type SOI ($\lambda_{so}=\Delta_{so}/3\sqrt{3}$), which opens nontrivial QSH bulk bandgap \cite{Kane05a, Kane05b} and induces topologically protected spin-filtered chiral edge states. The third term represents the staggered sublattice potential induced by the gate electric field ($E_v=\lambda_v/\alpha_v$ where $\alpha_v$ is the buckling-dependent parameter) which drives the QSH phase into a trivial insulating phase - termed here as topological switching. The fourth term is the spin-mixing Rashba SOI associated with the gate electric field \cite{Kane05a, Kane05b, Rashba09}, $\Delta_R=\alpha_RE_v$ where $\Delta_R=3\lambda_R/2$ and $\alpha_R$ is a Rashba SOI parameter. \par 

Here $c_{i\alpha}^\dagger (c_{i\alpha})$ is the creation (annihilation) electron operator with spin polarization $\alpha=\uparrow,\downarrow$ on site \textit{i}, the Pauli matrix $s^z$ describes the electron intrinsic spin while $s_{\alpha\beta}^z$ are the corresponding matrix elements describing the spin polarization $\alpha$ and $\beta$ on sites \textit{i} and \textit{j}, $v_{i}=+1(-1)$ for sublattice A (B), and $v_{ij}=\textbf{d}_{ik}\times\textbf{d}_{kj}=\pm1$ connects sites \textit{i} and \textit{j} on sublattice A (B) via the unique intermediate site \textit{k} on sublattice B (A). The nearest-neighbor bond vectors $\textbf{d}_{ik}$ and $\textbf{d}_{kj}$ connect the \textit{i} (\textit{k}) and \textit{k} (\textit{j}) sites on the A and B sublattices.\par

\begin{figure*}
\includegraphics[scale=0.7]{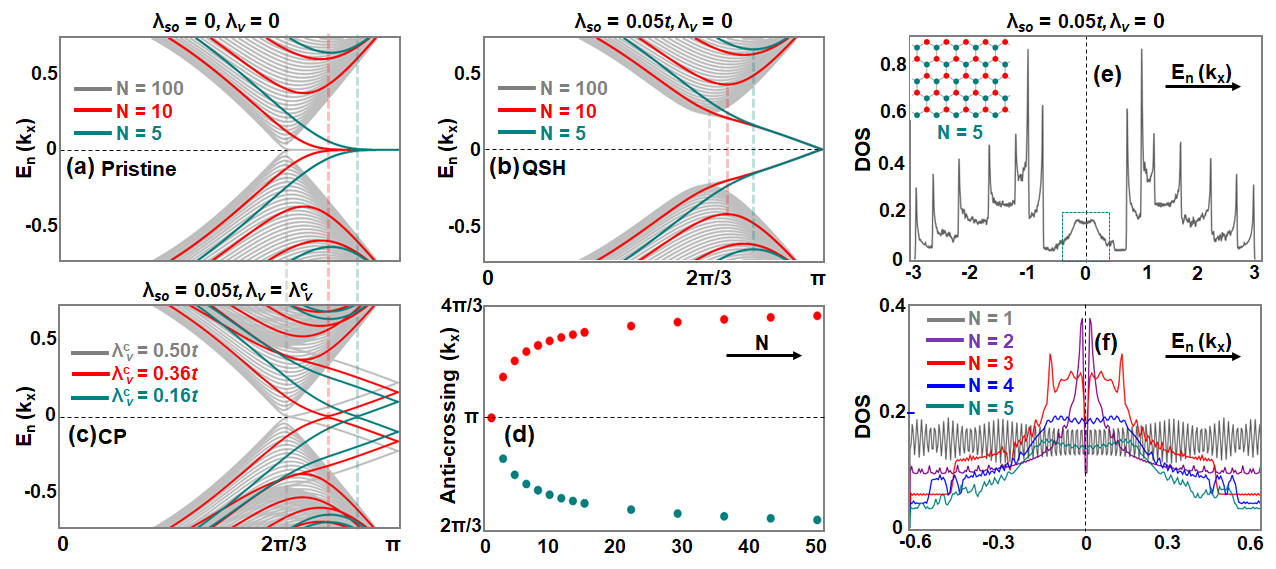}
\caption{\label{TPT-FSE}\textbf{Finite-size effects in quantum confined ZXNRs.} \textbf{(a-c)} Width dependence of one-dimensional electronic band dispersion for pristine ZXNRs hosting localized energy-zero edge states (a), spin-orbit coupled ZXNRs hosting QSH phase (b), and the gate induced critical point in spin-orbit coupled ZXNRs(c). In wide ZXNRs (N = 100) anti-crossing point lies at the valley ($k_x = 2\pi/3$) and the critical gate electric field reads $E_c = 2\Delta_{so}/\alpha_v$. In narrow quantum confined ZXNRs (N = 10, 5), anti-crossing point moves from valley towards TRIM $k_x=\pi$ and the critical gate electric field reduces from the SOI driven barrier, i.e., $E_c < 2\Delta_{so}/\alpha_v$. Moreover, around anti-crossing points, the energy spacing between edge states and the bulk subbands increases with decrease in width of ZXNRs. Such an increase in the bulk bandgap shows that topological switching is not accompanied by bulk bandgap closing and reopening in quantum confined ZXNRs. \textbf{(d)} Width dependence of momentum space location of anti-crossing point. \textbf{(e,f)} Density of states for ZXNRs in QSH phase with N = 5 (e) and edge state density of states for N = 1, 2, 3, 4 and 5 (f). Finite density of state around the energy-zero level shows that edge states in the QSH insulators with ZXNRs remain gapless even for ultra-narrow width. Here we set $a_0=t=1$ and $\lambda_R=0$.}
\end{figure*}

To begin with, by numerically diagonalizing the tight binding model, we study the finite-size effects on the pristine and spin-orbit coupled ZXNRs by varying the width of ZXNRs. For simplicity, the width of ZXNRs $W_z=\sqrt{3}Na_0/2$ is simulated by the number of zigzag chains (N) by setting $a_0=1$. In the pristine case, as shown in figure \ref{TPT-FSE} (a), the intrinsic topology of honeycomb lattice structures leads to strongly localized energy-zero flat edge states between valleys K and K$^\prime$, $\Delta k_x=K-K^\prime=2\pi/3a_0$, a nontrivial regime of the first Brillouin zone characterized by a non-vanishing winding number. The intrinsic SOI drives pristine ZXNRs into the QSH phase and disperses these energy-zero flat edge states into spin-filtered chiral edge modes, as shown in figure \ref{TPT-FSE} (b). Due to the presence of both time-reversal and inversion symmetry, the edge states are Kramers pairs, forming a fourfold degenerate Dirac point at the edge state crossing point, TRIM, $k_x=\pi$. Figure \ref{TPT-FSE} (c) shows that when a gate electric field is applied, the Kramers partners split along the energy axis while the twofold degenerate Dirac points in the spin down and spin up sectors move toward the corners of the nontrivial regime, valleys K and K$^\prime$ respectively. As a result, due to spin-valley locking, ZXNRs show a spin-polarized semi-metallic behavior at a critical point where $\lambda_v=\lambda_v^c$. When the strength of the staggered sublattice potential exceeds a threshold limit, the Dirac points in both spin sectors are gapped out at the anti-crossing points and the system enters the trivial regime.\par

The bulk and edge state electronic band dispersion, obtained via numerical diagonalization of the tight binding model, show a number of counter-intuitive features depending upon the width of ZXNRs, which may prove to be interesting for both fundamental and novel device applications in topological electronics.

\subsection{\label{1DTM}From 2D QSH insulator to 1D topological metal}
The trademark of spin-orbit coupled ZXNRs, i.e., spin-filtered chiral edge states in 2D QSH sheets, remains protected even when the system becomes effectively 1D and the QSH effect is no longer well defined. That is, as shown in figure \ref{TPT-FSE}(b), (i) the spin-filtered chiral edge states remain gapless even for ultra-narrow ribbons and (ii) the bulk bandgap increases with decrease in width. It implies that as one makes the ZXNR narrower, it retains its topological character, i.e., has well-defined 1D metallic modes associated with the edges, each with spin-momentum locking, and the bulk bandgap grows. So a narrow ZXNR remains a robust 1D topological metal, with a large energy separation between the edge states and the bulk subbands, characterized by a non-vanishing winding number associated with the intrinsic band topology. This non-intuitive result in ZXNRs seems interesting and differentiates ZXNRs from other 2D topological insulators with inverted band structure, where the effect of quantum confinement is to push the system toward a large-gap conventional insulator.\par

This observation can be understood from fundamental quantum mechanical considerations in narrow ZXNRs: sublattice-resolved intrinsic SOI, quantum confinement, and the longitudinal momentum-dependent inter-edge coupling. First of all, both the nontrivial bulk bandgap and the dispersing spin-filtered chiral edge states are indebted to the sublattice-resolved intrinsic SOI, i.e., next-nearest hopping localizes the bulk electrons while the electrons traversing along the edges remain effectively free. This SOI-induced mechanism remains true, irrespective of the ZXNR width. In addition, even in the 1D limit, as discussed below, the protection of spin-filtered chiral edge modes is guaranteed by the vanishing inter-edge coupling at TRIM $k_x=\pi$. On the other hand, the enhancement of a topological bulk bandgap is because of the quantum confinement effect on the bulk band spectrum. As shown in figure \ref{TQFET-FSE}(b), in the absence of gate potential, while bulk band varies as $E_{G/B}= |2\Delta_{so}|$ in the wide ZXNRs, the bulk bandgap varies as $E_{G/B}=|2\Delta_{so}+E_{qc}|$ in narrow ZXNRs. Here the energy gap $E_{qc}$ in the subband structure, induced by quantum confinement, is inversely proportional to the ZXNR width \cite{XNRs-Ezawa, XNRs-Kim, XNRs-Son, XNRs-Brey2, XNRs-Ezawa2, XNRs-Rubby}.

\subsection{\label{TS}Low-Voltage topological switching without bulk bandgap closing}
While finite-size effects have been widely studied in 2D-Xenes \cite{XNRs-Ezawa, XNRs-Kim, XNRs-Son, XNRs-Brey2, XNRs-Ezawa2, XNRs-Rubby}, effects of quantum confinement and momentum-dependent inter-edge overlapping on the gate-induced topological switching in spin-orbit coupled ZXNRs have received comparatively less attention. Similarly to the $\mathbb{PT}$-symmetric case, $\mathbb{PT}$-symmetry breaking via gate electric field also displays interesting features in narrow ZXNRs. As depicted in figure \ref{TQFET-FSE}(b) and \ref{TPT-FSE}(c), when the width of ZXNRs is smaller than a critical limit ($W_z^c$), (i) the critical gate electric field required for switching between gapless and gapped edge states decreases with decrease in width, and (ii) topological switching between gapless and gapped edge state spectrum is not accompanied by bulk bandgap closing and reopening.\par

First, with decreasing ZXNR width, the gate induced anti-crossing points in the edge state spectrum move away from the valleys toward TRIM. Since the momentum space location of anti-crossing points is directly associated with the threshold-voltage, the threshold-voltage decreases as the anti-crossing points move closer to the TRIM. For instance, in wide ZXNRs (N = 100), the spin-filtered Dirac points are gapped out exactly at the valleys K/K$^\prime$ and the SOI-induced barrier for critical electric field reads $E_c=2\Delta_{so}/\alpha_v$. On the other hand, in narrow quantum confined ZXNRs (N = 10, 5), the edge states are gapped out before reaching to the valleys K/K$^\prime$. As a result, the critical electric field reduces significantly from the SOI-driven barrier $E_c < 2\Delta_{so}/\alpha_v$ with decrease in the width. This trend suggests that the critical electric field has no lower bound and any nonzero electric field can open an energy gap in the edge state spectrum of ultra-narrow ZXNRs. \par

Second, the evolution of bulk subbands during topological switching from gapless to gapped edge state spectrum looks quite different for wide and narrow ribbons. In wide ZXNRs, during edge state evolution under the gate electric field, the bulk bandgap closes at the valleys when $E_c=2\Delta_{so}/\alpha_v$. At this point, the highest occupied molecular orbital and lowest unoccupied molecular orbital (HOMO-LUMO), carrying the same spin as that for edge states at particular valleys, of the bulk band spectrum become valley degenerate with the edge states. The bulk bandgap reopens when $E_c>2\Delta_{so}/\alpha_v$. It implies that the topological switching via electric field is accompanied by a quantum phase transition between $\mathbb{Z}_2$-nontrivial and $\mathbb{Z}_2$-trivial insulating phases where the bulk bandgap closes and reopens at the valleys. On the one hand, in narrow ZXNRs where bulk subbands and edge states are widely separated in energy due to quantum confinement, transitioning between gapless and gapped edge state spectrum occurs without bulk bandgap closing and reopening. The closing and reopening of the bulk bandgap is not necessary in narrow ZXNRs, as the 1D system is no longer a 2D topological insulator with a well-defined $\mathbb{Z}_2$ index. Hence, no bandgap closing and reopening is needed to switch the topology. \par

Such a finite-size-driven topological switching of edge state conductance, without bulk bandgap closing and reopening, is an entirely different concept from the previously studied quantum phase transition of the bulk band topology induced by symmetry breaking \cite{Ezawa13-TPT,Yang13,Rachel16, Matsumoto20,Schindler20}. Since the symmetry class of ZXNRs remains unchanged, irrespective of width, the quantum critical point for transitioning between $\mathbb{Z}_2$-nontrivial and $\mathbb{Z}_2$-trivial should remain the same, constrained by the SOI-induced barrier. However, apart from SOI terms, quantum confinement induces an extra contribution to the bulk bandgap of narrow ZXNRs. Since the gate electric field cannot manipulate the bandgap due to quantum confinement but the only one induced by SOI, it leads to topological switching of the edge state conductance via spin-filtered chiral modes without bulk bandgap closing. The critical electric field reads $E_c^{TS}<E_c^{QPT} = 2\Delta{so}/\alpha_v$, where the superscript "TS" and "QPT" represent topological switching and quantum phase transition respectively. It shows that the switching without bulk gap closing/reopening is a sheer consequence of the quantum confinement effect on the bulk band spectrum of narrow ZXNRs and can be verified from the calculated band dispersion (\ref{Ekx}). \par 

Accompanying finite-size effects on the edge state spectrum and quantum confinement effects on the bulk band spectrum, another critical phenomenon occurs, associated with the bulk band spectrum: the Rashba effect. For a specific width of ZXNRs, the Rashba SOI further lowers the critical gate electric field via topological quantum field effect on the bulk band spectrum \cite{Muhd-nano}. For quantum confined spin-orbit coupled ZXNRs, low-energy single-particle electronic dispersion in the vicinity of Dirac points reads as follows:
\begin{widetext}
\begin{equation}
E(k_x)= \pm \sqrt{v_{F}^2k_x^2+v_{F}^2k_n^2+\frac{1}{2}\Bigg| 6\sqrt{3}\lambda_{so}-\alpha_{v}E_v\Bigg(\frac{1}{2}+\sqrt{\frac{1}{4}+\Theta_{v(c)}\Bigg(\frac{2\alpha_R}{\alpha_v}\Bigg)^2}\Bigg)\Bigg|^2}\;. \label{Ekx}
\end{equation}
\end{widetext}
where $v_F=\sqrt{3}a_0t/2$ is the Fermi velocity, $\Theta_{v(c)}=1(0)$ for valence (conduction) bands in the QSH phase while $\Theta_{v(c)}=0(1)$ for valence (conduction) bands in the trivial phase and $k_n$ is the quantized transverse momentum along the confinement direction. Such finite-size-dependent quantization of transverse momentum divides the electronic band dispersion into infinite set of discrete subabands indexed by quantum number \textit{n} = 1, 2, 3.... Specific to our interest in this study, the band dispersion shows that quantum confinement induces an additional factor, $v_{F}^2k_n^2$, which enhances the bulk bandgap. The discretized transverse momentum $k_n$ is related to the longitudinal momentum $k_x$ as follows \cite{XNRs-Brey2}:
\begin{equation}
    k_x=\frac{k_n}{\tan(k_nW_z)}
\end{equation}

\subsection{\label{BT}Role of intrinsic topology in pristine ZXNRs}
The flat bands in the edge state spectrum of pristine ZXNRs are not generated from the intrinsic electronic spectrum of 2D-sheets but are rather indebted to the intrinsic band topology associated with the edge state wave functions. The electronic dispersion of pristine ZXNRs shows that a critical longitudinal momentum $k_x=k_x^c$ divides the momentum space regime for the extended (trivial) and the localized (nontrivial) edge states associated with gapped dispersing and gapless flat bands respectively. As shown in figure \ref{TPT-FSE}(a), the nontrivial regime of the first Brillouin zone hosting flat bands dwindles with decreases in the width of ZXNRs. That is, with decreasing width of the ZXNR, the location of the critical longitudinal momentum $k_x=k_x^c$ moves toward the TRIM $k_x=\pi$. As a result the critical longitudinal momentum reads $k_x=k_x^c>K$ for narrow ZXNRs, in contrast to $k_x=K$ in wide ZXNRs. \par

As depicted in figure \ref{TPT-FSE}(a) and \ref{TPT-FSE}(c), such finite-size effects on the pristine ZXNRs are intertwined with finite-size effects on the gate-induced topological switching in spin-orbit coupled ZXNRs. It is interesting to note that the momentum space location of gate-induced anti-crossing points in spin-orbit coupled ZXNRs is exactly the same as the critical longitudinal momentum $k_x=k_x^c$ in pristine ZXNRs. At this point the fourfold degenerate energy-zero flat bands in pristine ZXNRs are intrinsically gapped out by finite-size effects while the gate-induced twofold degenerate spin-filtered Dirac points in spin-orbit coupled ZXNRs are gapped out by the dominating gate electric field. It implies that the reduction of critical gate electric field in the spin-orbit coupled ZXNRs is intrinsically associated with the finite-size effects on the nontrivial character of pristine ZXNRs rather than mere manipulation of the intrinsic SOI. More specifically, the impact of intrinsic band topology of pristine ZXNRs on the electronic properties of spin-orbit coupled ZXNRs can be summarized as follows: in the nontrivial regime, while the critical momentum space location $k_x^c$ depends on the width of ZXNR, the strength of the critical gate electric field $E_c$ depends upon both the strength of SOI and the width of ZXNRs. This effect is further demonstrated by studying the real space wave functions for edge states, as shown below, that the reduction in the critical gate electric field is associated with gate-induced longitudinal momentum-dependence of inter-edge coupling in the vicinity of $k_x^c$.\par

\section{\label{TT}Topological edge state transport}
The existence and protection of spin-filtered chiral edge states in ultra-narrow ZXNRs and low-voltage topological switching from gapless to gapped edge states can be verified by studying the DOS, width and momentum dependence of inter-edge overlapping, and gate-induced inter-edge coupling for ZXNRs of various widths. 

\subsection{\label{DOS}Density of states}
The existence of spin-filtered chiral edge states in ultra-narrow ZXNRs is revisited by analyzing the DOS and the conductance quantization in ZXNRs. In the absence of a gate electric field, finite DOS at energy-zero represent the gapless edge states in the QSH phase, as shown in figure \ref{TQFET-FSE}(c), \ref{TPT-FSE}(e) and \ref{TPT-FSE}(f). Each edge state channel contributes $e^2/h$ to the conductance, leading to a quantized conductance of $2e^2/h$, figure \ref{TQFET-FSE}(d), which is an important signature of the QSH phase. The DOS remains finite at energy-zero level even for ultra-narrow ZXNRs, figure \ref{TPT-FSE}(e), a direct evidence of the existence of topological edge states in atomic-wide ZXNRs. When the gate electric field is switched on,the energy-zero DOS disappear in the $\mathbb{Z}_2$-trivial phase, figure \ref{TQFET-FSE}(c). Furthermore, a sharp disappearance of the DOS in the $\mathbb{Z}_2$-trivial phase shows that DOS measurement 
must be an efficient way for determining energy gap in edge spectra.\par

\subsection{\label{WMES}Width/Momentum dependence of edge states}
\begin{figure}
\includegraphics[scale=0.4]{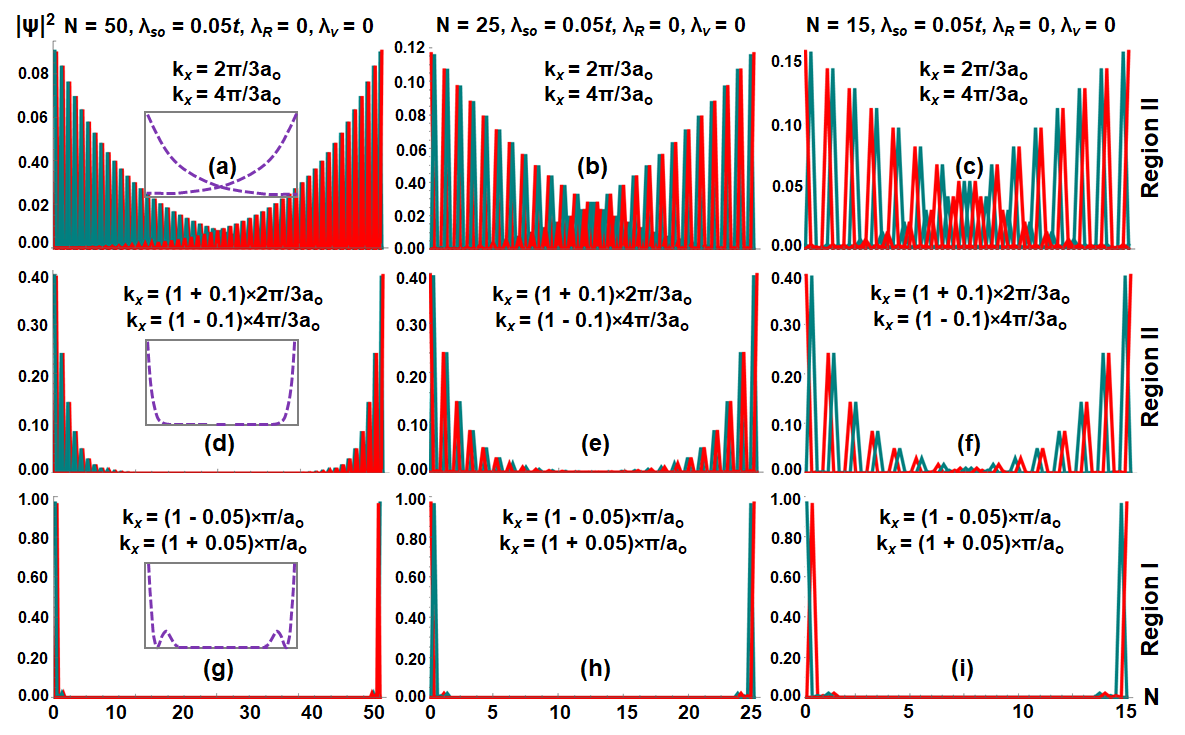}
\caption{\label{ES}\textbf{Longitudinal momentum and width dependence of real-space squared wave functions for the spin-filtered chiral edge states.} Real-space squared wave function for the spin-filtered chiral edge states near the Fermi energy of a ZXNRs with longitudinal momentum lying at Dirac/valley points $k_x=2\pi/3a_0$ \textbf{(a-c)}, away from Dirac points in the nontrivial regime \textbf{(d-f)}, and in the vicinity of TRIM $k_x \approx \pi/a_0$ \textbf{(g-i)}. The insets, dashed curves which are consistent with analytical tight binding model calculations, show that edge states are damped oscillatory in region I (g-i) while the edge states decay exponentially for momentum away from TRIM, in region II (a-f). For a fixed N, decay length of edge state decreases as $k_x$ moves from valleys towards the TRIM (from top to bottom). While there is no inter-edge overlapping in region I, finite inter-edge overlapping in region II increases with decrease in width (a-c). Here SOI parameters are taken as $\lambda_{so}/t= 0.05$ and $\lambda_R=0$ in the absence of gate potential $\lambda_v=0$. The horizontal axis is the confinement direction, along y-axis of the zigzag 2D-Xenes nanoribbon here.}
\end{figure}

In order to understand how the width and longitudinal momentum-dependence of inter-edge overlapping/coupling guarantees the protection of conducting edge states and assists electric field-driven topological switching, we investigate the real-space wave functions for edge states near the Fermi energy of a spin-orbit coupled ZXNR terminated at A and B sublattice sites respectively. As shown in figure \ref{TPT-FSE}, the spin-filtered chiral edge states are characterized by a range of longitudinal momentum $k_x\in(2\pi/3a_0,4\pi/3a_0)$, defining a nontrivial regime of the first Brillouin zone. In the vicinity of valleys $k_x \approx 2\pi/3a_0$ and $k_x \approx 4\pi/3a_0$, as depicted in figure \ref{ES}(a)-\ref{ES}(c), the real-space squared wave functions decay exponentially along the confined direction and have finite overlapping. With decrease in the width, though the amplitude near the edges increases, overlapping between edge states at the two sides of ZXNRs also increases. As one moves away from valleys toward TRIM $k_x=\pi$, due to large probability distribution near the edges, the amplitude of squared wave functions increases while the decay length decreases. For example, as shown in figure \ref{ES}(d)-\ref{ES}(f), nearly orthogonal squared wave functions indicate that the penetration depth of exponentially decaying edge states becomes much smaller than those around valleys K/K$^\prime$. Associated with longitudinal momentum around the TRIM $k_x=\pi$, as shown in figure \ref{ES}(g)-\ref{ES}(i), spin-filtered chiral edge states distributed near the edges appear to be completely orthogonal and, hence, the inter-edge overlap integral remains zero even for ultra narrow ZXNRs. \par

The accuracy of numerical tight binding results, describing the longitudinal momentum dependence of edge states in quantum confined ZXNRs, can be probed by obtaining explicit expressions for the wave functions for the spin-filtered chiral edge states by analytically solving tight binding model \cite{Mahdi-ZZ}. Based on the nature of wave functions, the edge state spectrum in ZXNRs can be divided into three regimes of momentum space: (i) In region I, in the vicinity of TRIM $k_x\approx\pi$ where the edge spectrum forms a fourfold degenerate Dirac point in the absence of gate electric field, the wave functions are damped oscillatory. (ii) In region II, $k_x\in[2\pi/3a_0,\pi)\cup(\pi,4\pi/3a_0]$ lying between the Dirac points, the wave functions at the edges decay exponentially along the confined direction. (iii) In region III, away from the Dirac point $2\pi/3a_0>k_x>4\pi/3a_0$, the wave functions are oscillatory in nature and represent the localized 1D edge states due to admixing with bulk subbands. It implies that the spin-filtered chiral edge states are formed by a combination of wave functions in region I and II and the nature of these wave functions changes from exponentially decaying to damped oscillatory as $k_x$ moves from region II to region I. It shows that numerical tight binding calculations are consistent with analytical tight binding results in region I and II as shown in the insets of figure \ref{ES}. \par

While tight binding dispersion and DOS confirms the existence of gapless edge states in atom-wide ZXNRs, the analysis of width and momentum dependence of edge state wave functions leads to the following three important outcomes: (i) protection of 1D topological metal, (ii) gate-induced tunability of inter-edge coupling via correspondence between various momentum space regimes and real space edge termination, and (iii) size-dependent optimization of topological switching.

\subsection{\label{TP-1DTM}Protection of 1D topological metal}
The classification of edge state longitudinal momentum $k_x$ shows that the edge states in ZXNRs are similar to those in a conventional quantum Hall strip where translational symmetry is preserved along the strip. The spin-filtered chiral edge states along the two sides of ZXNRs are associated with different $k_x$, and they do not hybridize with each other even when there is a finite overlap along the confined direction in region II \cite{Edges-Halperin,Edges-MacDonald}. On the other hand, in region I where the energy and momentum of edge states around the crossing point $k_x=\pi$ are nearly equal and they can possibly couple to open an energy gap, their wave functions do not overlap in a finite space. It implies that, in the absence of gate electric field, spin-filtered chiral edge states do not hybridize/couple even in ultra-narrow ZXNRs.\par 

While excellent conductance quantization, important signature in many topological states, and its robustness are well-known features of the quantum Hall effect, even the extensively studied 2D topological insulators \cite{Bernevig06,Konig07,Liu-typeII, Knez-TypeII, Du-TypeII,Konig08} with inverted band structure show experimentally much more fragile conductance quantization at low temperatures \cite{Konig07,Knez-TypeII, Du-TypeII}. So a question arises: is there any advantage to QSH effect in ZXRNs or will the topological protection remain relatively fragile? Within the accuracy of electronic dispersion, DOS, quantized conductance, and momentum-dependence of edge state wave functions found via numerical tight binding approximations, the topological protection of QSH states in ZXNRs is equivalent to that for quantum Hall insulators. It suggests that the edge states in ZXNRs are far more stable than other topological insulator materials with inverted band structure.\par

As mentioned above, the answer lies in the energy and momentum space location of conducting modes on opposite edges in a finite-size geometry. The existence of different momentum-space locations for the edge state crossings and anti-crossings in ZXNRs is highly contrasting from other 2D topological insulator materials with inverted band structure, in which edge state crossing and anti-crossing points coexist, and in which hybridization due to inter-edge overlapping opens an energy gap and leads to a gapped edge state spectrum \cite{Zhou08}. Furthermore, it is also explicitly demonstrated in section-\ref{eeC} that edge states in honeycomb structures remain protected against electron-electron Coulomb interaction which may become inevitable due to inter-edge overlapping in narrow ribbons. In short, we are not aware of any experimental obstacles that may cause potential threat to edge state conductance quantization in ZXNRs. However, precise control of the zigzag edge is required in device fabrication.\par

\subsection{\label{GIC}Momentum dependence of gate-induced inter-edge coupling}
The resemblance of damped oscillatory behavior around $k_x=\pi$ to the one in spin-orbit coupled armchair 2D-Xenes nanoribbons (AXNR) \cite{Mahdi-AC} suggests that the dynamical evolution of edge states remains independent of particular edge termination in region I. On the other hand, the exponentially decaying wave functions in region II are directly associated with the particular edge termination on A and B sublattice sites and, hence, their penetration depth can be tuned via gate-induced staggered sublattice potentials.\par

\begin{figure}
\includegraphics[scale=0.4]{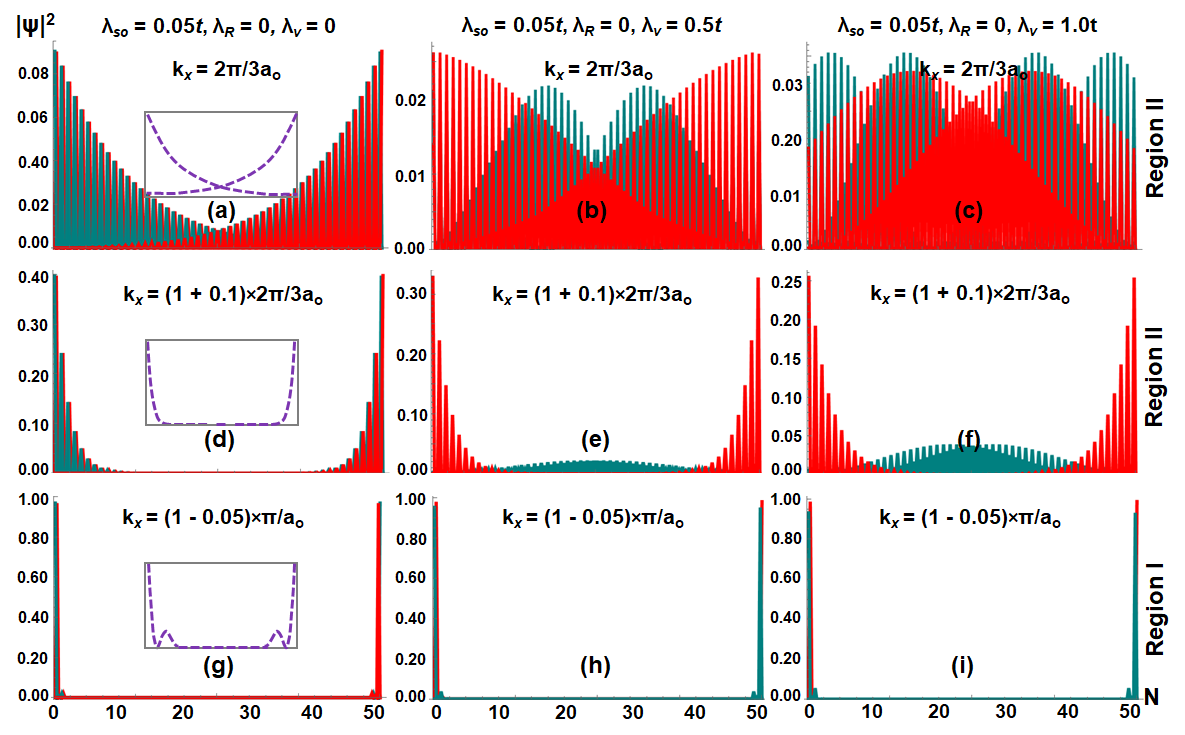}
\caption{\label{ES-Ez}\textbf{Longitudinal momentum dependence  of gate-induced inter-edge coupling.} For a fixed width of N=50, effect of gate electric field on real-space squared wave function for the spin-filtered chiral edge states associated with longitudinal momentum $k_x$ lying in region II \textbf{(a-f)} and region I \textbf{(g-i)}. In the vicinity of valleys $k_x=2\pi/3a_0$ (a-c), critical gate electric field localizes both the spin-down and spin-up sectors by turning exponentially decaying edge states into sinusoidal form. Such gate-induced enhancement in the penetration depth of chiral edge states and the gate-induced inter-edge coupling leads to an energy gap in the edge state spectrum when gate electric field exceeds critical limit. As one moves away from Dirac point, $k_x\approx5\pi/6a_0$ (d-f), similar evolution occurs in spin-down sector but the spin-up sector remains always exponentially decaying and traversing. In the vicinity of TRIM $k_x=\pi/a_0$ (g-i), penetration depth of edge states remains insensitive to gate electric field effect and both the spin up and the spin down chiral edge states remains traversing along the edges even for quite large gate electric field. Here SOI parameters are taken as $\lambda_{so}/t= 0.05$ and $\lambda_R=0$. The horizontal axis is the confinement direction, along y-axis of the nanoribbon here.}
\end{figure}

To further investigate the dependence of the gate electric field effect on the longitudinal momentum $k_x$, we study the gate electric field modulation of edge state wave functions associated with various longitudinal momenta. As shown in figure \ref{ES-Ez}(g)-\ref{ES-Ez}(i), our numerical calculations show that the damped oscillatory edge states in region I remain insensitive to the gate electric field, i.e., penetration depth remains the same and both the spin up and down edge states remain damped oscillatory and traversing along the edges even for very large gate electric field. On the other hand, as shown in figure \ref{ES-Ez}(a)-\ref{ES-Ez}(f), exponentially decaying edge states in region II are highly sensitive to gate electric field effects. First of all, in the vicinity of momentum $k_x \approx 5\pi/6a_0$ as shown in figure \ref{ES-Ez}(d)-\ref{ES-Ez}(f), gate electric field hybridizes spin-down edge states while the amplitude of spin-up edge states decreases with increasing electric field but they remain uncoupled. It implies that, spin up edge states remain exponentially decaying and traversing while spin down edge states become sinusoidal and gapped in this regime of longitudinal momentum, consistent with tight binding electronic dispersion. As one moves toward the valley $k_x \approx 2\pi/3a_0$ as shown in figure \ref{ES-Ez}(a)-\ref{ES-Ez}(c), it can be clearly seen that gate electric field induces coupling between overlapping exponentially decaying wave functions in both spin up and down sectors and localizes them. The period of these exponentially turned sinusoidal wave functions decreases with increase in the gate voltage. Similar gate-controlled edge state dynamics appears on the other valley, $k_x \approx 4\pi/3a_0$, but the spin character is interchanged due to electric field-induced spin-valley locking. \par

Based on the edge state dynamics, we came to the following conclusion: in the absence of gate electric field, the non-vanishing overlap between edge sates in region II do not hybridize/couple to open energy gap as they lie at different longitudinal momenta. However, mainly due to spin-valley locking, the gate electric field splits fourfold degenerate Dirac point in region I and moves the spin-polarized twofold Dirac points toward region II. The finite inter-edge overlapping in region II allows the gate electric field to induce coupling between spin-filtered inter-edge states and open an energy gap in the edge state spectrum.\par

\subsection{\label{opt}Size-dependent optimization of topological switching}
In region II as shown in figure \ref{ES}(a)-\ref{ES}(f), an increase in the overlap between inter-edge states with decreases in width indicates the enhancement of gate-induced inter-edge coupling in narrow ZXNRs. The gate electric field utilizes such enhancement of the penetration depth and, hence, inter-edge overlapping in region II to lower the critical gate electric field required for topological switching between gapless and gapped edge states. The consistency of both numerical and analytical tight binding results indicates that the finite-size effect assistance in topological switching is an artifact of this gate-induced momentum-dependent coupling between wave functions along the edges. \par

\begin{figure}
\includegraphics[scale=0.73]{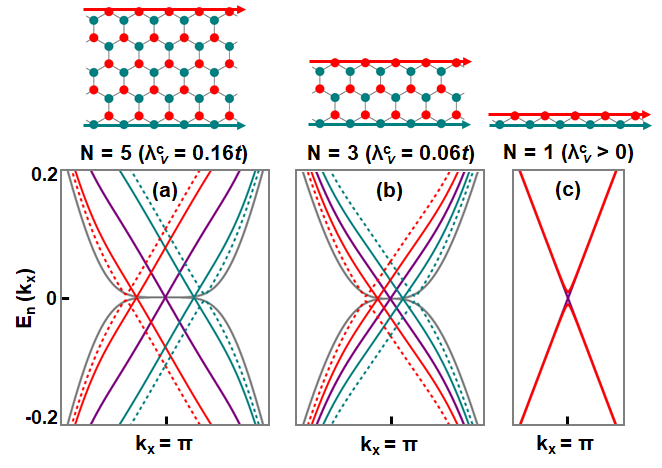}
\caption{\label{TPT-N135}\textbf{Size-dependent threshold-voltage.} Topological switching from gapless to gapped edge states for a ZXNR with N = 5 (a), N = 3 (b), and N = 1 (c). Here grey lines represent edge states for pristine ZXNR, purple lines represent gapless edge states with $\lambda_{so}=0.05t$ and $\lambda_v=0$, red and cyan solid lines represent critical phases with $\lambda_v=\lambda_v^c$, and red and cyan dashed lines represent gapless edge states with $\lambda_v=0.22t$ (a), $\lambda_v=0.12t$ (b), and $\lambda_v=0.02t$ (c). For N = 1, any positive threshold sublattice potential $\lambda_v^c$ opens energy gap in the edge state spectrum. Here we set $a_0=t=1$, $\lambda_{so}=0.05t$, and $\lambda_R=0$.}
\end{figure}

The width dependence of gate-induced inter-edge coupling is also consistent with the width dependence of tight binding electronic dispersion: with a decrease in the width, the gate-induced anti-crossing points move away from valleys toward TRIM and, thus, the threshold-voltage decreases. Furthermore, figure \ref{TPT-N135} shows that there is no fundamental limit on the threshold-voltage: For a single zigzag chain, N = 1, the edge states of both pristine and spin-orbit coupled ZXNR form a gapless Dirac dispersion where crossing and anti-crossing points coexist at $k_x=\pi$. As a result, any non-zero value of staggered sublattice potential opens an energy gap in the edge state spectrum. 

\section{\label{eeC}Effect of electron-electron Coulomb interactions}
Furthermore, though the electron-electron Coulomb interactions become inevitable in quantum confined ZXNRs, spin-filtered chiral edge conducting channels may remain gapless even when both intra- and inter-edge Coulomb interactions are present. For example, contrary to pristine ZXNRs where Coulomb interactions lead to energy gap by lifting the fourfold degeneracy of energy-zero flat bands in the edge state spectrum \cite{XNRs-Son,ZGNR-RG,GNR-Fujita,GNR-Yang}, it has been explicitly shown that the QSH phase in spin-orbit coupled ZXNRs remain stabilized against intra-edge Coulomb interactions \cite{Xu-Moore-06}. Moreover, the mass term produced by backward scattering, which may have originated from the mixing of right and left moving chiral modes carrying the same spin polarization and located at opposite edges of finite-size ZXNRs, can be suppressed in the large SOI limit and, hence, the spin-filtered chiral edge states may also remain protected against unscreened inter-edge Coulomb interactions \cite{Mahdi-ZZ}. It can be justified by a simple argument based on the interplay between the strength of intrinsic SOI and the screening length of Coulomb interactions: Since the decay length of spin-filtered chiral edge states in ZXNRs is inversely proportional to the intrinsic SOI, the overlaps between oppositely moving spin-filtered chiral modes are suppressed with increasing SOI. It shows that, in the large SOI limit, ultra-narrow ZXNRs can be described by the tight binding model where both intra- and inter-edge Coulomb interactions are effectively absent. Thus, even in the presence of Coulomb interactions, a large SOI limit renders gapless spin-filtered chiral edge states since the reduced inter-edge overlap diminishes the backward scattering terms. \par

Based on a similar argument, our findings provide another ground: In the QSH phase, spin-filtered chiral states associated with a longitudinal momentum $k_x=\pi$ in region I remain protected against backward scattering due to vanishing inter-edge overlap. On the other hand, in the presence of a gate electric field, unscreened inter-edge Coulomb interactions may also assist topological switching by inducing an energy gap due to finite inter-edge coupling between edge states associated with longitudinal momentum $k_x$ lying in region II of the Brillouin zone, similar to the finite-size effect. In passing, unlike the finite-size effect, which is characterized by critical longitudinal momentum $k_x^c$ and remains the same in both pristine and spin-orbit coupled ZXNRs, the effect of Coulomb interactions in pristine ZXNRs is completely different from spin-orbit coupled ZXNRs.

\section{\label{TQD}Low-voltage topological quantum devices}
The analogy between the rich momentum-dependent behavior of edge states and the gate-controlled inter-edge coupling in the spin-orbit coupled ZXNRs leads to the following two phenomena which are critical for topological quantum devices: (i) In the absence of gate electric field and hence Rashba SOI, spin-filtered chiral edge edge states with fourfold degenerate Dirac point at TRIM $k_x=\pi$ remain gapless even for ultra-narrow ZXNRs. Vanishing inter-edge coupling across the crossing point guarantees that the spin-filtered chiral edge states (enabling dissipationless and quantized conductance) remain topologically protected against backscattering and hence the deviation from conductance quantization - a figure of merit in QSH materials. (ii) Since the gate electric field splits fourfold Dirac point at TRIM and moves spin-filtered twofold Dirac points toward valleys, gate-induced coupling due to finite overlap between spin-filtered inter-edge states across anti-crossing points assists in opening the energy gap in the spin-filtered chiral edge states and lowers the critical gate electric field. This artifact of the finite-size effect, dwindled nontrivial regime of the Brillouin zone without affecting the bulk band topology and reduced critical gate electric field without affecting the quantized edge state conductance in the QSH phase, provides an ideal platform for devising energy-efficient low-voltage topological quantum devices. \par

\begin{figure}
\includegraphics[scale=0.8]{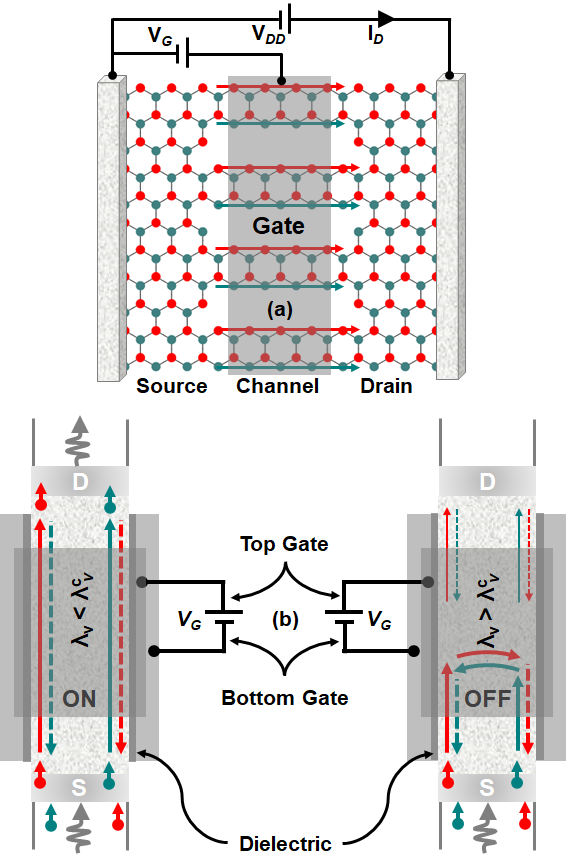}
\caption{\label{TQFET}\textbf{Topological quantum field effect transistor.} \textbf{(a)} Schematic representation for a TQFET configuring multiple quantum confined ZXNRs allowing conduction between source and drain. \textbf{(b)} Topological switching driven by gate electric field which tunes a ZXNR form on-state with gapless edge modes ($\lambda_v<\lambda_v^c$) to off-state with gapped edge modes ($\lambda_v>\lambda_v^c$). Here V\textsubscript{G} is the gate-voltage, V\textsubscript{DD} is the supply-voltage, and I\textsubscript{D} is the source-to-drain current.}
\end{figure}

To exemplify the advantages of utilizing spin-orbit coupled ZXNRs for computing technologies, we explicitly demonstrate the working principle of a topological quantum field effect transistor (TQFET) and compare its critical functionalities with a MOSFET. Unlike a MOSFET, where a conventional semiconductor is utilized as a channel material and conduction is enabled via bulk electronic states, TQFET configures a topological insulator material as a channel in which the dissipationless current is carried by topologically protected edge modes. In a blueprint for a TQFET, the gate electric field tunes a topological insulator material from a topological insulating phase (on-state) to a conventional insulating phase (off-state), a phenomenon known as topological switching. In other words, the topological switching mechanism relies on transitioning between gapless (on-state) and gapped (off-state) edge modes. Such a gate electric field-driven topological switching is a fundamentally different mechanism compared to a traditional carrier inversion in conventional semiconducting switching devices. Figure \ref{TQFET} shows a schematic representation for a TQFET configuring quantum confined ZXNR as a channel between the source and drain. First, the existence of 1D gapless edge states in the ultra-narrow ZXNRs promises the availability of large edge state conducting modes for enhanced signal-to-noise ratio via multiple edge state channels, figure \ref{TQFET}(a). It allows optimized geometry for a TQFET where an array of ZXNRs, set apart by trivial insulating layers/wires along vertical/lateral direction, is sandwiched between top and bottom gates separated by top and bottom dielectrics. \par

Second, the reduction in threshold-voltage with decrease in the channel width, even though the topological bulk bandgap increases, overturns a general wisdom of utilizing narrow gap and wide channel materials for reducing threshold-voltage in a standard field effect transistor analysis. For example, in a blueprint topological transistor where topological switching is implemented via bulk bandgap closing and reopening, materials with large bulk bandgap require an unrealistically large threshold-voltage \cite{Molle-natmat17, Pseudo-TFET-HgTe}. Moreover, TQFET with quantum confined ZXNRs is in high contrast to MOSFET in which width-dependence of the threshold-voltage V\textsubscript{th} depends upon the isolation technique used for transistor fabrication \cite{tsividis2011}: the effective threshold-voltage in a narrow channel device increases with decreases in width when the transistor is made using the LOCOS (local oxidation of silicon) process while decreases with decrease in width when the transistor is made in a shallow-trench-isolation (STI) process. That is, unlike the size dependence of threshold-voltage on isolation technique in MOSFET, the reduction of threshold-voltage in a TQFET is an intrinsic property of ZXNRs associated with topological and quantum mechanical functionalities. It suggests that, along with vastly different conduction and switching mechanisms, the technological aspects required for fabricating a TQFET with ZXNRs also radically differ from those of MOSFETs: There is no fundamental requirement of specialized technological/isolation techniques for a low-voltage TQFET with an energy-efficient switching mechanism. \par   

Third, the reduction in threshold-voltage becomes important for reducing the supply voltage (V\textsubscript{DD}) in a low-voltage switching device if the subthreshold swing is compressible. That is, power dissipation \cite{Ionescu11} P $\approx$ I\textsubscript{OFF}V\textsubscript{DD}$^3$ can be reduced while maintaining the device performance (I\textsubscript{ON}) by simultaneous scaling down $V_{th}$ and $V_{DD}$ and, thus, keeping the overdrive constant ($\propto$(V\textsubscript{DD} - V\textsubscript{th})$^2$). In a MOSFET, incompressible subthreshold swing leads to an exponential increase in I\textsubscript{OFF} in the transfer characteristics \cite{Ionescu11}. That is, for every 60 mV at room temperature, there is more than tenfold increase in I\textsubscript{OFF}. In contrast to MOSFETs, this is not a problem in a TQFET where subthreshold swing can be tuned via topological quantum field effect \cite{Muhd-nano}, a combined effect of electric field and tunable Rashba SOI that allows overcoming the “Boltzmann’s tyranny”. Power dissipation can be lowered by reducing the threshold-voltage via geometric optimization of quantum confined ribbons of QSH materials while keeping subthreshold swing subthermal via strain engineering, buckling parameterization, tuning inter-orbital hopping, and normalization of intrinsic atomic SOI. In summary, quantum confined ZXNRs with optimized geometry may prompt the progress of topological computing technologies with vastly lower energy consumed per operation than CMOS technologies and greet the Moore's trajectory of transistor miniaturization, doubling transistors per chip and doubling the processing power every two years.\par

\section{\label{TBM}Experimental realization and device fabrication}
Though a full experimental exploitation of 2D-Xenes is yet to be explored for device applications, mainly due to several challenges imposed by restricted synthesis methodologies, substrate effects, and stability issues, an experimental confirmation of the electronic properties and the buckled structure for 2D-Xenes has been realized through angle-resolved photoemission spectroscopy (ARPES) and scanning tunneling microscopy (STM), respectively. Furthermore, the epitaxial growth of both group-IV \cite{Molle-natmat17} and group-V \cite{Karim-GV} 2D-Xenes has been realized on different substrates. For instance, the atomic arrangement of group-IV buckled 2D-Xenes has been detailed through epitaxy of silicene on metallic (Ag(111), Ir(111), and ZBr$_2$) \cite{Si-Lalmi,Si-Vogt,Si-Lin,Si-Feng, Si-Chiappe1,Si-Fleurence,Si-Meng} and semiconducting (MoS$_2$) \cite{Si-Chiappe2} substrates, STM studies of germanene on metallic (Au(111), Pt(111), Al(111) and Hex-AIN) \cite{Ge-Dvila,Ge-Li,Ge-Bampoulis,Ge-Derivaz,Ge-DAcapito} and semiconducting (MoS$_2$) \cite{Ge-Zhang} substrates, and epitaxial stanene on Bi$_2$Te$_3$ substrates \cite{Sn-Zhu}. The epitaxial synthesis has also been extended to group-V 2D-Xenes, for instance, a monolayer of phosphorene on Au(111) substrate \cite{P-Zhang} showing silicene-like semiconducting character.\par

While the tight binding model with Kane-Mele type SOI describes a hypothetical freestanding ZXNRs with poor chemical stability, a substrate supporting the epitaxial synthesis is highly desired for real-world applications. However, the supporting substrate brings other challenges into play due to concurrent bonding interactions. For instance, a metallic substrate short-circuits the edge states of interest and destroys the topological protection. On the other hand, semiconducting MoS$_2$ substrate can stabilize 2D-Xenes with protected edge states; however, 2D-Xenes render into a metal due to compressive strain \cite{Si-Chiappe2,Ge-Zhang}. Similar problems persist for stanene on Bi$_2$Te$_3$ substrate \cite{Sn-Zhu}. \par

Recently, it has been shown that epitaxially deposited bismuthene on the insulating silicon carbide substrate SiC(0001) is a large bandgap QSH insulator where structural and electronic properties have been confirmed by STM and ARPES measurements \cite{Reis17}. However, as compared to freestanding buckled Bi(111) bilayers, considerably larger lattice constant of 5.35 \text{\AA} stabilizes Bi/SiC into an energetically favorable planar honeycomb configuration \cite{Hsu15}. With all its interesting aspects, the planar honeycomb configuration is not desirable for gate-induced topological switching. Furthermore, it is predicted that both As and Sb are plagued by similar problems \cite{Li18}.\par

This experimental odyssey of 2D-Xenes promises that the fabrication of topological devices is just a step away. The possible first experimental step to corroborate our main prediction for device integration, scientific studies, and technological applications is the synthesis of buckled ZXNRs with protected edge states on a weakly interacting semiconducting substrate. In this direction, the growth of functionalized 2D-Xene sheets on a suitable semiconducting substrate would be a promising development that can bring an obvious benefit for the realization of low-voltage topological devices \cite{Muhd-nano}. For instance, functionalized bismuth monolayers BiX \cite{Song14} and Bi$_2$XY \cite{Zhou18} where X/Y = H, F, Cl, and Br stabilize with quasi-planar/low-buckled structure and the strong on-site SOI opens the topological bandgap at the Dirac points formed by low-lying $p_{x}$ and $p_{y}$ orbitals. Furthermore, recently corroborated first-principle calculations show that gate-controlled topological quantum phase transition between different topological states can be realized in functionalized bismuth monolayer \cite{Igor21}.\par

\section{\label{Con}Conclusion}
It is demonstrated that, in a finite-size geometry, ZXNRs display unique physical characteristics associated with their intrinsic band topology and the finite-size effects such as longitudinal momentum-dependent inter-edge overlapping between spin-filtered chiral edge states and the quantum confinement effect on the bulk band spectrum. While the damped oscillatory modes around the edge state crossing momentum remain completely orthogonal and guarantee protected spin-filtered chiral edge states even in ultra-narrow ribbons, enhanced gate-induced inter-edge coupling between exponentially decaying edge states around the anti-crossing points reduces the gate electric field required for topological switching between gapless and gapped edge states. In addition, quantum confinement enhances the SOI-induced bandgap in the nontrivial phase and leads to topological switching without bulk bandgap closing. On the one hand, it reduces the threshold-voltage by lowering the SOI-induced barrier in the bulk; on the other hand, it enhances the bulk bandgap even in lighter monoelemental 2D-Xenes such that the detrimental contributions from the bulk material to the edge current are avoided and allows safe residing of the chemical potential within this gap. Furthermore, similar to wide ZXNRs, the Rashba effect enhances the bandgap in the trivial phase. Hence, a large nontrivial bulk bandgap by quantum confinement effect to decouple the conducting edge states from bulk subbands and a large trivial bandgap by the Rashba effect to overcome thermal excitation makes quantum confined narrow ZXNRs ideal for engineering energy-efficient low-voltage topological quantum devices.\par

The proposed mechanism for optimizing topological switching and devising concepts for topological electronics is applicable to all 2D-Xene sheets ranging from silicene to bismuthene as well as other 2D topological insulators with honeycomb lattice structure. In principle, the threshold-voltage depends upon the momentum space location of anti-crossing points in the edge state spectrum, which is the same for both pristine and spin-orbit coupled ZXNRs. Quantitatively, the threshold value depends upon both the strength of intrinsic SOI and the width of ZXNRs. In addition, the width of ZXNRs, $W_z=\sqrt{3}Na_0/2$, depends upon both the number of zigzag lines $N$ and the lattice constant $a_0$. Increasing lattice constants, ranging from 2.46 \text{\AA} for graphene to 5.35 \text{\AA} for bismuthene \cite{Reis17}, suggest that the critical width ($W_z^c$) would be different for different 2D topological insulator sheets, even if the number of zigzag lines $N$ is fixed. Furthermore, a wide range of the intrinsic SOI, from 0.00057 meV for graphene to 435 meV for bismuthene \cite{Reis17}, indicates that the threshold-voltage would be different for different 2D topological insulators sheets, even if the width of the ribbon is fixed. \par

Considering the wide applicability and generality of the proposed mechanism, we presented a generalized mechanism that clearly indicates how tight binding electronic dispersion, DOS, and the penetration depth of edge state wave functions and, thus, the threshold-voltage depend on the number of zigzag chains. In wide ZXNRs, with $W_z>W_z^c$, the SOI ($\Delta_{so}$) induced barrier imposes a limit on the threshold voltage ($\lambda_v^c=2\Delta_{so}$), which could be unrealistically large for 2D-Xenes. However, when $W_z<W_z^c$, the threshold-voltage decreases with decrease in width ($\lambda_v^c<2\Delta_{so}$). A qualitative width dependence study shows that the threshold-voltage can be lowered without any fundamental limit. For instance, in ultra-narrow ribbons, N=1 say, any non-zero electric field can switch the edge state conductance by opening an energy gap in the edge state spectrum. Considering a large variation in the strength of intrinsic SOI and size of lattice constants for 2D-Xenes, we did not mention any single threshold limit as a figure of merit for a topological transistor. Rather, we highlighted that a topological transistor is more flexible to tunability of critical parameters than its conventional counterpart, MOSFET. More defined engineering heads-up toward the manufacture of an ideal topological transistor can be extracted for a specific 2D-Xene configured as a channel material.\par

Similar to the gate voltage, finite-size effects can also be employed to tune the exchange interaction in 2D magnetic topological insulators \cite{Ezawa1,Ezawa2,Ezawa3,Hogl-G,Li-MnPX,Liang_ABO,Zhou18,Igor21,Babar19,Nadeem20} and optical probing of topological signatures in 2D materials \cite{Igor20}. For example, the critical regime in an antiferromagnetic topological insulator can be optimized to design a topological spin transistor via gate-induced topological switching of edge state spin transport. This study may also be generalized to study other topological phases such as topological superconductors \cite{TSC-San,TSC-Ezawa}. In a finite-size geometry, Majorana bound states localized along the edges of 2D topological superconductors \cite{TSC-PotterLee} can be decoupled from bulk states for robust information processing. \par

\begin{acknowledgments}
This research is supported by the Australian Research Council (ARC) Centre of Excellence in Future Low-Energy Electronics Technologies (FLEET Project No. CE170100039), Australian Research Council (ARC) Professional Future Fellowship (FT130100778), and funded by the Australian Government.
\end{acknowledgments}



\nocite{*}
\bibliography{aipsamp}
\end{document}